\documentclass[11pt]{article}
\usepackage{arxiv}
\DeclareUnicodeCharacter{00A7}{\S}
\DeclareUnicodeCharacter{00B0}{\textdegree}
\DeclareUnicodeCharacter{00B1}{\ensuremath{\pm}}
\DeclareUnicodeCharacter{00D7}{\ensuremath{\times}}
\DeclareUnicodeCharacter{0394}{\ensuremath{\Delta}}
\DeclareUnicodeCharacter{03C3}{\ensuremath{\sigma}}
\DeclareUnicodeCharacter{2026}{\ldots}
\DeclareUnicodeCharacter{2192}{\ensuremath{\rightarrow}}
\DeclareUnicodeCharacter{2212}{\ensuremath{-}}
\DeclareUnicodeCharacter{2248}{\ensuremath{\approx}}
\DeclareUnicodeCharacter{2265}{\ensuremath{\ge}}
\usepackage[table]{xcolor}
\definecolor{TietTableStripe}{HTML}{F3F6FA}
\definecolor{TietTableRule}{HTML}{CBD5E1}
\usepackage{color}
\usepackage{fancyvrb}

\DefineVerbatimEnvironment{Highlighting}{Verbatim}{commandchars=\\\{\}}
\usepackage{framed}
\definecolor{shadecolor}{RGB}{241,243,245}
\newenvironment{Shaded}{\begin{snugshade}}{\end{snugshade}}

\newcommand{\BuiltInTok}[1]{\textcolor[rgb]{0.00,0.23,0.31}{#1}}

\newcommand{\CommentTok}[1]{\textcolor[rgb]{0.37,0.37,0.37}{#1}}

\newcommand{\ControlFlowTok}[1]{\textcolor[rgb]{0.00,0.23,0.31}{\textbf{#1}}}

\newcommand{\DecValTok}[1]{\textcolor[rgb]{0.68,0.00,0.00}{#1}}

\newcommand{\ImportTok}[1]{\textcolor[rgb]{0.00,0.46,0.62}{#1}}

\newcommand{\KeywordTok}[1]{\textcolor[rgb]{0.00,0.23,0.31}{\textbf{#1}}}
\newcommand{\NormalTok}[1]{\textcolor[rgb]{0.00,0.23,0.31}{#1}}
\newcommand{\OperatorTok}[1]{\textcolor[rgb]{0.37,0.37,0.37}{#1}}

\newcommand{\SpecialCharTok}[1]{\textcolor[rgb]{0.37,0.37,0.37}{#1}}
\newcommand{\SpecialStringTok}[1]{\textcolor[rgb]{0.13,0.47,0.30}{#1}}
\newcommand{\StringTok}[1]{\textcolor[rgb]{0.13,0.47,0.30}{#1}}

\RecustomVerbatimEnvironment{Highlighting}{Verbatim}{commandchars=\\\{\},fontsize=\small}
\makeatletter
\@ifpackageloaded{caption}{}{\usepackage{caption}}
\AtBeginDocument{%
\ifdefined\contentsname
  \renewcommand*\contentsname{Table of contents}
\else
  \newcommand\contentsname{Table of contents}
\fi
\ifdefined\listfigurename
  \renewcommand*\listfigurename{List of Figures}
\else
  \newcommand\listfigurename{List of Figures}
\fi
\ifdefined\listtablename
  \renewcommand*\listtablename{List of Tables}
\else
  \newcommand\listtablename{List of Tables}
\fi
\ifdefined\figurename
  \renewcommand*\figurename{Figure}
\else
  \newcommand\figurename{Figure}
\fi
\ifdefined\tablename
  \renewcommand*\tablename{Table}
\else
  \newcommand\tablename{Table}
\fi
}
\@ifpackageloaded{float}{}{\usepackage{float}}
\floatstyle{ruled}
\@ifundefined{c@chapter}{\newfloat{codelisting}{h}{lop}}{\newfloat{codelisting}{h}{lop}[chapter]}
\floatname{codelisting}{Listing}

\makeatother
\makeatletter
\@ifpackageloaded{caption}{}{\usepackage{caption}}
\@ifpackageloaded{subcaption}{}{\usepackage{subcaption}}
\makeatother

\title{PySynthea: A Python-Native Framework for Scalable Synthetic
Healthcare Data Generation}

\author{%
Roberto Cruz%
\thanks{\texttt{roberto.cruz@tiet.ai}}%
 \qquad %
David Rey-Blanco%
\thanks{\texttt{david.rey@tiet.ai}}%
\\[0.4em]
\small TietAI%
}

\date{2026-05-21}

\begin{document}
\maketitle

\begin{abstract}
\noindent Synthetic healthcare data is increasingly important for
research, education, and machine learning development where access to
real patient data is limited by privacy and governance constraints.
While Synthea provides a widely adopted framework for generating
realistic longitudinal electronic health record data, its current
implementation presents adoption barriers for many researchers and data
scientists due to deployment complexity and limited integration with
modern Python-based workflows.

This paper introduces PySynthea, a Python-native reimplementation of
Synthea designed to improve accessibility, extensibility, and
interoperability within the scientific Python ecosystem. The framework
provides modular synthetic patient generation, configurable healthcare
simulation pipelines, and support for standard healthcare data formats
while integrating naturally with tools such as pandas and machine
learning workflows. By reducing operational complexity and aligning
synthetic data generation with the dominant data science ecosystem,
PySynthea aims to accelerate experimentation and broaden the use of
synthetic healthcare data in research and applied AI development.
\end{abstract}

\bigskip

\section{Introduction}\label{sec-introduction}

The development of artificial intelligence and machine learning systems
for healthcare has accelerated rapidly over the last decade, with
growing evidence that data-driven models can support clinical decision
making, operational planning, population health analytics, and
biomedical discovery
\citep{rajkomar2018scalable, esteva2019guide, topol2019high}. The
training, evaluation, and benchmarking of these systems depend
critically on access to large, longitudinal, and clinically realistic
electronic health record (EHR) data. In practice, however, real EHR data
is constrained by privacy regulations, institutional review processes,
contractual restrictions, and risks of re-identification, all of which
limit the speed at which experimentation can occur and reproducibility
can be achieved \citep{el2015anonymising, benitez2010beyond}.

Synthetic healthcare data has emerged as a complementary resource that
can mitigate these barriers without disclosing information about real
individuals. High-quality synthetic EHR datasets enable several
activities that would otherwise be difficult or impossible at scale,
including reproducibility of clinical informatics studies, benchmarking
of machine learning models across institutions, education of clinical
and informatics trainees, end-to-end testing of healthcare software and
integration pipelines, and the safe pre-training and prototyping of
machine learning models prior to their deployment on protected health
information \citep{chen2021synthetic, gonzales2023synthetic}. As
synthetic data has matured, it has become a recognized component of
modern healthcare data engineering and a vehicle for reproducible
research in clinical artificial intelligence.

Despite well-developed regulatory frameworks, access to real EHR data
remains operationally complex. The Health Insurance Portability and
Accountability Act (HIPAA) in the United States \citep{hipaa1996} and
the General Data Protection Regulation (GDPR) in the European Union
\citep{gdpr2016} place substantive constraints on how protected health
information may be collected, processed, shared, and retained.
Institutional approvals, data use agreements, and ethical review
processes routinely add months or years to project timelines, and the
resulting datasets are typically constrained to a specific cohort, time
window, or analytical purpose. Even when de-identified datasets are
released, re-identification risks remain non-trivial, especially when
records can be linked to auxiliary information
\citep{benitez2010beyond}. The combination of legal, organizational, and
statistical constraints means that real EHR data is often unavailable
for many of the early, exploratory stages of machine learning research.

Synthetic data does not eliminate these concerns entirely, but it does
change the operational geometry of the problem. A synthetic dataset that
is generated from a defined statistical model rather than from real
individuals can be shared across institutions, embedded in tutorials,
included in software test suites, or used to validate large data
pipelines without disclosing protected information. This makes synthetic
generation an enabling technology for both education and reproducible
research in healthcare informatics.

Although healthcare informatics historically grew up around enterprise
Java and C\# stacks, the broader landscape of data analytics and machine
learning has consolidated around Python. Numerical and tabular
computation in Python is supported by NumPy \citep{harris2020array} and
pandas \citep{mckinney2010pandas}; deep learning is dominated by PyTorch
\citep{paszke2019pytorch} and TensorFlow \citep{abadi2016tensorflow};
classical machine learning is widely practiced through scikit-learn
\citep{pedregosa2011scikit}; large-scale data processing uses Dask
\citep{rocklin2015dask} and Apache Spark \citep{zaharia2016spark}
through PySpark; and the day-to-day environment for many researchers is
the Jupyter notebook \citep{kluyver2016jupyter}. Foundation models for
clinical text and tabular EHR data are also typically distributed in
Python-friendly frameworks such as the Hugging Face Transformers library
\citep{wolf2020transformers, wornow2023foundation}. The result is a
strongly Python-first research ecosystem in which most experimentation,
model development, and pipeline orchestration is expected to happen.

In this ecosystem, the cost of bridging into a non-Python tool is
non-trivial. It introduces additional moving parts in continuous
integration, complicates reproducibility, increases the cognitive load
on researchers and students, and creates impedance between data
generation and downstream model development. For synthetic healthcare
data to fully serve the Python-based ML ecosystem, the data generators
themselves need to live within it.

This paper introduces PySynthea, a Python-native reimplementation of the
Synthea synthetic patient generation framework
\citep{walonoski2018synthea}. PySynthea preserves the conceptual
foundations of Synthea --- modular, state-machine-based disease modules;
longitudinal simulation of patients from birth to death; demographically
grounded population sampling; multi-format export including FHIR ---
while making the generator first-class within the Python ecosystem.

The contributions of this work are as follows. First, we describe a
Python-native implementation that is installable via standard tooling
such as \texttt{pip} and \texttt{uv} \citep{uv_astral}, requires no JVM,
and exposes a clean Python API alongside a \texttt{click}-based
command-line interface. Second, we describe the architectural decisions
that make the system modular and extensible: a state-machine engine that
loads the original Synthea Generic Module Framework (GMF) JSON modules
unchanged; a separation between simulation engine, world model, and
exporters; and a configuration layer that mirrors Synthea's properties
model. Third, we describe interoperability features oriented toward the
modern Python data ecosystem, including pandas-compatible exports, FHIR
R4 bundles, and design patterns for streaming generation. Fourth, we
discuss notebook-friendly workflows, parallel and batched generation,
and integration points for downstream machine learning pipelines.
Finally, we discuss tradeoffs, limitations, and a research agenda that
builds on the Python-native foundation toward GPU acceleration,
LLM-augmented module authoring, and reinforcement-learning environments
grounded in synthetic clinical trajectories.

The remainder of the paper is organized as follows. Section 2 reviews
the original Synthea project and its data model. Section 3 examines the
limitations of existing synthetic data generation workflows from the
perspective of a Python-first user. Section 4 articulates the design
goals of PySynthea. Sections 5 and 6 describe the system architecture
and the synthetic patient generation pipeline. Section 7 details Python
ecosystem integration. Section 8 covers healthcare data standards and
export formats. Section 9 discusses performance and scalability. Section
10 presents example workflows and use cases. Section 11 discusses
tradeoffs, Section 12 outlines future work, and Section 13 concludes.

\section{Background: Synthea}\label{sec-background}

Synthea is an open-source synthetic patient population simulator
originally developed by The MITRE Corporation
\citep{walonoski2018synthea, synthea_github}. The project was motivated
by the recognition that many healthcare informatics activities ---
interoperability testing, software validation, education, prototyping,
and benchmarking --- do not require real patient data, but do require
data that is structurally and statistically realistic. Synthea fills
this gap by generating fully synthetic patients together with
longitudinal clinical histories that mirror the structure of real
electronic health records.

A Synthea generation run produces a population of synthetic individuals
along with their associated entities. These typically include
demographic attributes (such as age, sex, race, ethnicity, and
geographic location), encounters (ambulatory visits, inpatient
admissions, emergency visits), conditions, medications, procedures,
observations and vital signs, immunizations, care plans, allergies,
devices, and imaging studies. Each patient has a complete simulated life
trajectory from birth (or initialization) to death or to the end of the
simulation horizon, and the resulting record includes both timing and
clinical content for every modeled event.

At the core of Synthea is the Generic Module Framework (GMF), a
state-machine formalism for encoding disease progression and care
pathways. Each module is a directed graph of states --- including
initial states, simple states, delays, guards, encounters, condition
onsets and ends, medication orders, procedures, observations, care plan
starts and ends, symptoms, and terminal states --- connected by direct,
conditional, or probabilistically distributed transitions
\citep{walonoski2018synthea, dube2021framework}. At each simulation
timestep, a patient's active modules are evaluated, transitions are
sampled, and clinical events are emitted into the patient's record.

This formalism has several important properties. It separates clinical
content from execution semantics, so domain experts can author or modify
modules without touching the simulation engine. It is probabilistic,
supporting realistic variability across simulated patients. It is
composable, since modules can invoke submodules and reference each other
through shared patient attributes. And it is auditable, since the
resulting trajectories can be traced through the underlying state graph.

Around this state-machine core, Synthea layers a population model that
draws demographics from real census data, a healthcare system model that
includes providers and payers, cost data calibrated against publicly
available pricing references, and a set of clinical reference resources
such as CDC growth charts, immunization schedules, and biometric
correlations.

\subsection{Interoperability and
Standards}\label{interoperability-and-standards}

A major contributor to Synthea's adoption is its support for healthcare
interoperability standards. The Fast Healthcare Interoperability
Resources (FHIR) standard \citep{bender2013hl7, hl7_fhir_r4} provides a
modern, RESTful, JSON- or XML-based representation of clinical
resources, and is now the dominant standard for exchanging healthcare
data. Synthea natively exports synthetic patients and their records as
FHIR bundles, in addition to flat CSV exports, CCDA documents, and other
formats. Generated resources use standardized coding systems including
SNOMED CT \citep{snomed_ct} and LOINC \citep{loinc} for diagnoses and
observations.

The combination of realistic longitudinal records and
standards-conformant export formats has made Synthea a default tool for
validating FHIR servers, training SMART-on-FHIR applications
\citep{mandel2016smart}, and prototyping healthcare apps without
exposing real patient information.

\subsection{Current Use Cases}\label{current-use-cases}

Synthea is used today across several distinct communities. In education,
it provides a safe and reproducible substrate for teaching clinical
informatics, FHIR, and healthcare data science. In machine learning
research and prototyping, it enables benchmarking of pipelines and
models against datasets that mirror the structural complexity of real
EHRs while remaining fully shareable
\citep{chen2021synthetic, gonzales2023synthetic, yan2022multifaceted}.
In healthcare software development, Synthea is widely used for
end-to-end testing of EHR integrations, data engineering pipelines, and
interoperability gateways. During the COVID-19 pandemic, Synthea was
extended to model COVID-19-specific trajectories and to provide
synthetic datasets useful for early research before real data could be
made widely available \citep{walonoski2020covid}.

These use cases share a common pattern: they require synthetic data that
is rich, longitudinal, and standards-conformant, and they value the
absence of regulatory constraints over absolute statistical fidelity.
PySynthea is designed to preserve these characteristics while making the
tool itself more accessible.

\section{Limitations of Existing Synthetic Data Generation
Workflows}\label{sec-limitations}

This section discusses the practical limitations encountered when using
existing synthetic data generation tooling --- and Synthea in particular
--- from the perspective of users whose workflows are centered on the
Python ecosystem. These limitations should be read as friction relative
to modern data science practice rather than as criticism of Synthea
itself, whose contributions to the field are substantial.

\subsection{Deployment Complexity}\label{deployment-complexity}

The reference implementation of Synthea is written in Java and is
typically operated either via a build system (Gradle), a runnable JAR,
or a Docker container. While these mechanisms are robust and well
established, they introduce real operational friction in environments
that are otherwise Python-centric. A researcher who simply wants to
obtain 1,000 synthetic patients in a Jupyter notebook is typically
required to install a Java Development Kit, clone a separate repository,
invoke an external build command, navigate an output directory of CSV or
FHIR files, and then re-ingest the resulting files into pandas or
another Python tool \citep{walonoski2018synthea}. Each of these steps
adds latency, surface area for configuration errors, and difficulty for
users unfamiliar with the JVM toolchain.

For students, applied researchers, and software engineers whose default
workflow is \texttt{pip\ install} followed by \texttt{import}, this
multi-runtime setup is a meaningful adoption barrier. By contrast, a
Python-native generator can be installed with a single command, imported
as a library, and invoked directly from the same notebook in which
downstream analysis lives.

\subsection{Limited Integration with Python
Workflows}\label{limited-integration-with-python-workflows}

Beyond deployment, a deeper integration cost arises when
Synthea-generated data is consumed by a Python-based analytics or ML
pipeline. The natural unit of data interchange in this context is the
pandas \texttt{DataFrame} or, in larger settings, the Dask or PySpark
distributed equivalent. Synthea's outputs, however, must first be
materialized to disk in CSV, JSON, or FHIR bundle format, then re-parsed
and re-shaped before they can be used. This round-trip introduces
serialization overhead, fragile parsing logic for nested FHIR
structures, and an explicit gap between generation and analysis.

Modern workflows also place a high premium on interactive iteration:
regenerating a small cohort with slightly different parameters in a
notebook, inspecting it, and iterating again. A non-Python generator
that must be invoked through an external process and whose outputs must
be parsed back into the host language is inherently misaligned with this
pattern.

\subsection{Barriers to Rapid
Experimentation}\label{barriers-to-rapid-experimentation}

Many of the most interesting use cases for synthetic data are inherently
iterative. A researcher exploring how an ML model's performance varies
with disease prevalence may want to regenerate populations many times
with different parameter values. A developer building a custom disease
module may want to make a change, regenerate a small cohort, inspect the
resulting records, and iterate within seconds. An educator may want to
weave generation into a lecture or a hands-on lab without asking
students to set up an external runtime.

These workflows favor frameworks in which generation parameters live as
plain Python objects, generation can be invoked in a single function
call, and outputs are immediately available as in-memory objects. They
are less compatible with workflows that require a build system, a
separate process, and file-based round-trips.

\subsection{Ecosystem Fragmentation}\label{ecosystem-fragmentation}

A consequence of these structural mismatches is a degree of
fragmentation between the healthcare simulation ecosystem and the
broader ML ecosystem. Many of the most important downstream tools ---
pandas \citep{mckinney2010pandas}, NumPy \citep{harris2020array},
PyTorch \citep{paszke2019pytorch}, scikit-learn
\citep{pedregosa2011scikit}, Dask \citep{rocklin2015dask}, PySpark
\citep{zaharia2016spark}, Airflow \citep{apache_airflow}, and Jupyter
\citep{kluyver2016jupyter} --- live in a single connected graph of
imports, configuration files, and packaging conventions. A Java-based
generator sits outside this graph and must be bridged through file
formats and shell invocations, rather than through library calls and
shared in-memory data structures.

The cost of this fragmentation falls disproportionately on the users who
are best positioned to benefit from synthetic data: applied ML
researchers, students, and small healthcare AI teams that lack dedicated
infrastructure engineers.

\subsection{Extensibility Challenges}\label{extensibility-challenges}

Finally, although Synthea's modules are themselves JSON files that are
in principle language-agnostic, modifying the simulation engine itself
--- for example, to add new state types, alter timestep semantics,
change how attributes propagate, or instrument the engine for tracing
--- requires familiarity with the Java codebase, its build system, and
its testing infrastructure. For researchers whose primary expertise is
in machine learning and clinical data science rather than enterprise
Java engineering, this is a steep on-ramp.

A Python-native engine that mirrors the original semantics can lower
this barrier substantially. Researchers can extend the simulation engine
using the same language they use for analysis, debug it with familiar
tools, and contribute improvements through the same workflow they
already use for their machine learning code.

\section{Design Goals of PySynthea}\label{sec-design-goals}

PySynthea is designed around a small set of explicit goals that follow
directly from the limitations described in the previous section. They
are summarized in Table~\ref{tbl-design-goals} and discussed below.

\begingroup
\small
\renewcommand{\arraystretch}{1.5}

\begin{longtable}[]{@{}
  >{\centering\arraybackslash}p{(\linewidth - 2\tabcolsep) * \real{0.1500}}
  >{\raggedright\arraybackslash}p{(\linewidth - 2\tabcolsep) * \real{0.8500}}@{}}
\caption{Design goals of
PySynthea.}\label{tbl-design-goals}\tabularnewline
\toprule\noalign{}
\begin{minipage}[b]{\linewidth}\centering
Goal
\end{minipage} & \begin{minipage}[b]{\linewidth}\raggedright
Description
\end{minipage} \\
\midrule\noalign{}
\endfirsthead
\toprule\noalign{}
\begin{minipage}[b]{\linewidth}\centering
Goal
\end{minipage} & \begin{minipage}[b]{\linewidth}\raggedright
Description
\end{minipage} \\
\midrule\noalign{}
\endhead
\bottomrule\noalign{}
\endlastfoot
Accessibility & Install with \texttt{pip\ install} or \texttt{uv\ add};
no external runtime; usable from the first line of a notebook. \\
Interoperability & First-class integration with pandas, NumPy, PyTorch,
Dask, and other tools in the Python data ecosystem. \\
Reproducibility & Deterministic generation conditional on a seed;
explicit configuration; ability to reproduce a cohort exactly across
runs. \\
Extensibility & Modular simulation engine with documented extension
points for new state types, exporters, and modules. \\
Scalability & Multi-threaded and process-based generation, streaming
export, and design hooks for distributed and GPU-accelerated futures. \\
Standards Fidelity & Preserve compatibility with the original Synthea
GMF JSON modules and export FHIR R4 bundles, CSV, and JSON. \\
Notebook-First & Designed for interactive use: small cohorts return
quickly, generation can be inspected and re-run interactively. \\
\end{longtable}

\endgroup

\emph{Accessibility} is the central goal: removing the JVM dependency
and shipping a pure-Python package eliminates an entire class of
installation problems and brings synthetic data generation within reach
of every user who is already running Python. \emph{Interoperability}
with the dominant data ecosystem ensures that generated data flows
directly into downstream analysis and modeling without unnecessary
intermediate steps. \emph{Reproducibility} is a non-negotiable
requirement for any tool used in scientific research and is supported by
deterministic seeding and explicit configuration.

\emph{Extensibility} recognizes that synthetic data is rarely useful in
only its default form: users routinely need new disease modules, custom
exporters, or modifications to the underlying engine, and these tasks
must be tractable in the user's own language. \emph{Scalability}
acknowledges that synthetic data generation is sometimes a small
in-notebook task and sometimes a multi-million-patient batch job, and
the framework must accommodate both ends of this spectrum.
\emph{Standards fidelity} ensures that PySynthea continues to operate
within the broader Synthea ecosystem and produces data that is
compatible with downstream tools that already expect Synthea-style
outputs. Finally, the \emph{notebook-first} orientation reflects the
practical reality that the daily working environment for most users in
this space is the Jupyter notebook \citep{kluyver2016jupyter}.

\section{System Architecture}\label{sec-architecture}

PySynthea is organized into four primary packages within a single
installable distribution. The \texttt{engine} package implements the
simulation engine, including the module loader, state machine semantics,
transition logic, condition evaluation, and the top-level generator. The
\texttt{world} package contains the domain model: persons, health
records, demographics, geographic locations, providers, and payers. The
\texttt{export} package implements exporters for FHIR R4 and other
formats. The \texttt{helpers} package provides configuration utilities
and common functionality. A separate \texttt{resources} tree, packaged
with the distribution, holds the disease modules, provider and payer
data, geographic and demographic data, cost tables, and clinical
reference resources that mirror those of upstream Synthea.

\begin{figure}

\centering{

\pandocbounded{\includegraphics[keepaspectratio]{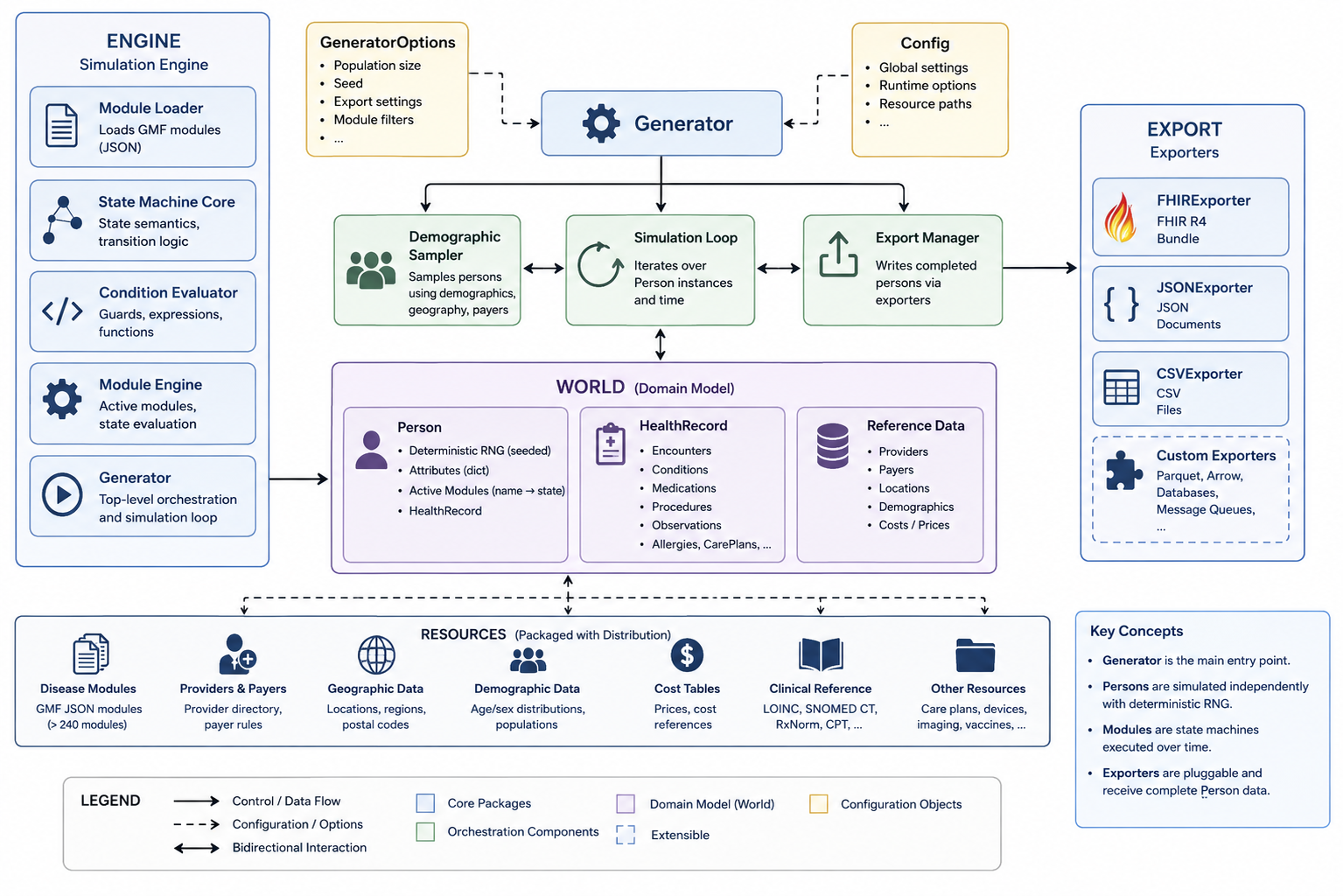}}

}

\caption{\label{fig-architecture}High-level system architecture of
PySynthea. The \texttt{Generator} orchestrates the simulation loop;
\texttt{Person} instances carry attributes and a health record; the
\texttt{Module} engine drives state-machine evaluation over time; and
the export layer produces FHIR, JSON, and CSV outputs.}

\end{figure}

The high-level relationship between these components is shown in
Figure~\ref{fig-architecture}. The \texttt{Generator} object is the main
entry point. It is configured by a \texttt{GeneratorOptions} object and
a \texttt{Config} object, and it coordinates the demographic sampler,
the simulation loop over individual \texttt{Person} instances, the
module engine, and the exporters. Each \texttt{Person} carries a
deterministic random generator seeded from the global seed, a dictionary
of attributes, a set of active modules with their current states, and a
\texttt{HealthRecord} that accumulates clinical events.

The simulation engine is built around three central abstractions. A
\texttt{Module} represents a single state machine, identified by name
and composed of a set of named \texttt{State} instances and a designated
initial state. A \texttt{State} is a typed node in the state machine ---
for example, an \texttt{Encounter}, a \texttt{ConditionOnset}, a
\texttt{MedicationOrder}, a \texttt{Delay}, or a \texttt{Terminal} ---
that executes side effects on the patient when entered and exposes a
transition policy. A \texttt{Transition} selects the next state given
the current patient context. The set of supported state types in
PySynthea mirrors that of upstream Synthea and includes initial, simple,
delay, guard, set-attribute, counter, encounter and encounter-end,
condition-onset and condition-end, allergy-onset and allergy-end,
medication order and end, care-plan start and end, procedure, vital
sign, observation, multi-observation, diagnostic report, symptom, death,
call-submodule, device and device-end, supply-list, imaging-study,
vaccine, and physiology states.

Modules are loaded at startup from JSON files using the existing Synthea
GMF schema. This was an explicit architectural decision: rather than
re-author hundreds of disease modules in a new format, PySynthea parses
and executes the established GMF JSON directly, which preserves more
than 240 disease modules and the long history of clinical content
embedded in them. From the user's perspective, the same module that runs
in upstream Synthea also runs in PySynthea, simplifying validation and
enabling incremental contributions to the shared module library.

The exporter layer is structured around a \texttt{PatientExporter}
interface, with concrete implementations such as \texttt{FHIRExporter}
producing FHIR R4 bundles. Each exporter receives a fully simulated
\texttt{Person} and writes its contents to disk or to a downstream sink.
The architecture allows new exporters --- for example, Parquet, Arrow,
or direct database writers --- to be added without modifying the
simulation engine.

\section{Synthetic Patient Generation Pipeline}\label{sec-pipeline}

The synthetic patient generation pipeline in PySynthea proceeds through
six conceptual stages. At each stage, the design preserves the semantics
of upstream Synthea while exposing Python-native interfaces that are
convenient for downstream consumers.

\subsection{Population Initialization}\label{population-initialization}

When the \texttt{Generator} is constructed, it loads the configuration,
sets the global random seed, instantiates the demographic sampler over
the configured geographic scope, loads providers and payers, and loads
the set of active modules. The population size, age range, sex filter,
and reference date are taken from the \texttt{GeneratorOptions} object.
A typical invocation looks as follows.

\begin{Shaded}
\begin{Highlighting}[]
\ImportTok{from}\NormalTok{ synthea }\ImportTok{import}\NormalTok{ Generator, GeneratorOptions}

\NormalTok{options }\OperatorTok{=}\NormalTok{ GeneratorOptions()}
\NormalTok{options.population\_size }\OperatorTok{=} \DecValTok{1000}
\NormalTok{options.state }\OperatorTok{=} \StringTok{"California"}
\NormalTok{options.city }\OperatorTok{=} \StringTok{"San Francisco"}
\NormalTok{options.seed }\OperatorTok{=} \DecValTok{12345}
\NormalTok{options.threads }\OperatorTok{=} \DecValTok{4}

\NormalTok{generator }\OperatorTok{=}\NormalTok{ Generator(options)}
\NormalTok{generator.run()}
\end{Highlighting}
\end{Shaded}

\subsection{Demographic Sampling}\label{demographic-sampling}

For each requested patient, the demographic sampler produces an initial
set of attributes --- sex, race, ethnicity, geographic location, ZIP
code, and birth date --- drawn from real census-derived distributions.
The sampler is deterministic when seeded, ensuring that the same set of
patients can be regenerated across runs. Social determinants of health,
such as income, education, and housing-related variables, can be sampled
from accompanying tables \citep{walonoski2018synthea}.

\subsection{Disease Progression}\label{disease-progression}

Once a patient has been initialized, the engine instantiates the active
modules for that patient and begins the main simulation loop. Time
advances in discrete steps --- by default one week, configurable through
the properties file --- from the patient's birth date to either their
simulated death or the reference date, whichever comes first. At each
timestep, each module is evaluated: the engine inspects the patient's
current state in that module, runs the state, samples a transition, and
advances the state pointer if the transition completes within the
current timestep. Modules can declare delays, which suspend further
progression until a specified amount of simulated time has elapsed, and
guards, which block progression until a specified condition becomes
true.

\subsection{Clinical Event Simulation}\label{clinical-event-simulation}

The states executed during disease progression emit clinical events into
the patient's \texttt{HealthRecord}. An \texttt{Encounter} state opens a
new encounter, attaching subsequent condition onsets, medication orders,
procedures, and observations until the corresponding
\texttt{EncounterEnd}. Each event is associated with standardized codes
(SNOMED CT for conditions \citep{snomed_ct}, LOINC for observations
\citep{loinc}, RxNorm-style codes for medications, and so on) drawn from
the module definitions. Cost information is attached to encounters,
procedures, medications, immunizations, labs, devices, and supplies
using the cost tables shipped with the distribution.

\subsection{Temporal Sequencing}\label{temporal-sequencing}

A central property of Synthea-style data is its longitudinal nature:
events are not isolated, but ordered in time and causally related.
PySynthea preserves this through its timestep-based engine. A patient
who acquires diabetes early in life may, over the course of decades of
simulated time, accumulate medications, laboratory observations,
complications, and additional encounters in a clinically plausible
order. The same temporal coherence applies across modules --- for
example, pregnancy modules interacting with prenatal care encounters and
immunization schedules.

\subsection{Export Generation}\label{export-generation}

\begin{figure}

\centering{

\pandocbounded{\includegraphics[keepaspectratio]{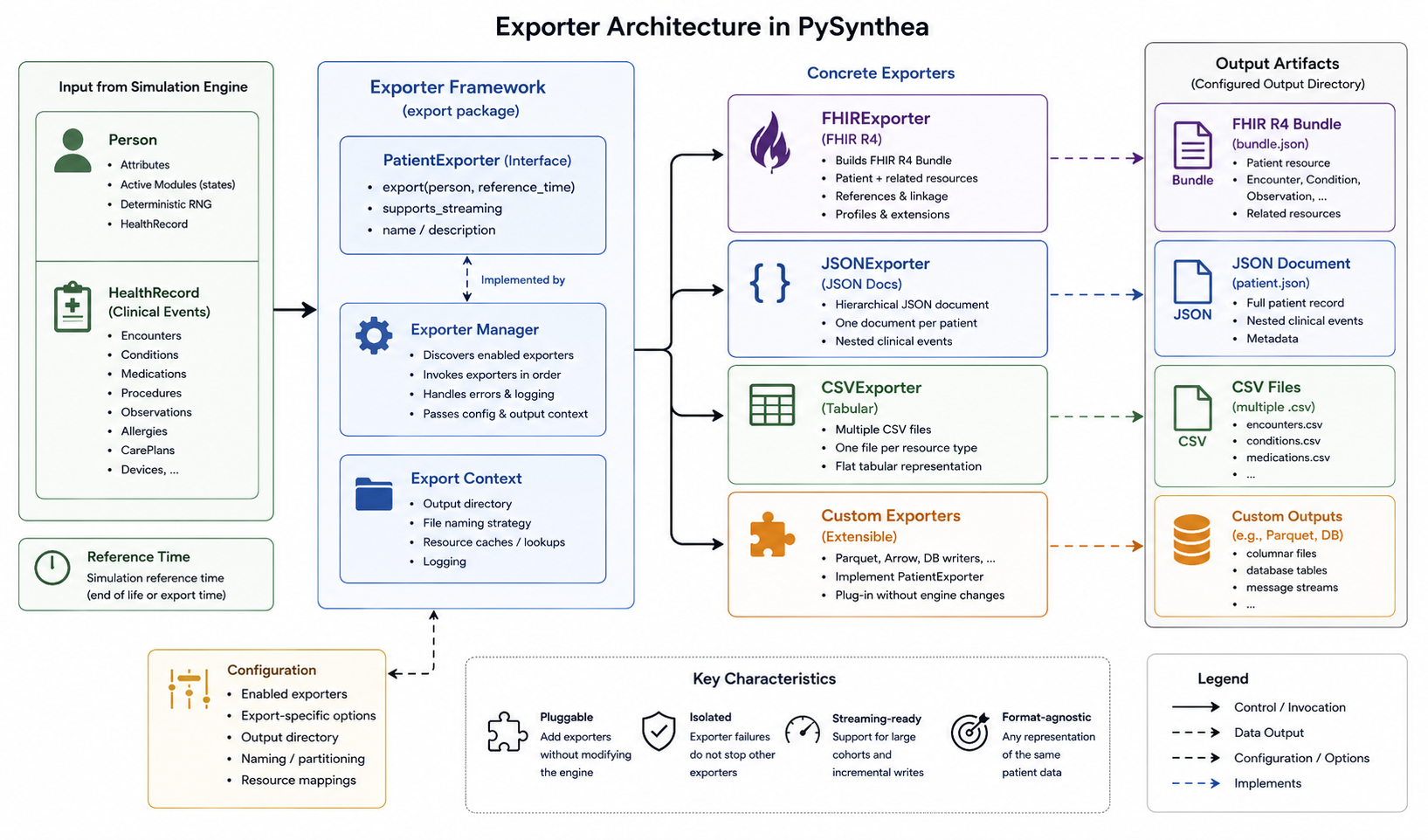}}

}

\caption{\label{fig-pipeline}End-to-end synthetic patient generation
pipeline. Each patient is initialized, demographically sampled,
simulated through the module engine across a longitudinal timeline, and
finally exported in one or more formats.}

\end{figure}

Once a patient's life has been simulated, each enabled exporter is
invoked with the patient and the simulation reference time. The
\texttt{FHIRExporter} constructs a FHIR R4 bundle containing the patient
resource and each associated event as a separate FHIR resource, while
CSV and JSON exporters produce flat tabular and document-oriented
outputs respectively. Exporters write to a configured output directory
and can be enabled or disabled independently. A schematic of the full
pipeline is shown in Figure~\ref{fig-pipeline}.

\section{Python Ecosystem Integration}\label{sec-python-integration}

The defining design choice of PySynthea is to be a first-class member of
the Python ecosystem. This section describes the integration patterns
that follow from that choice.

\subsection{Native DataFrame Access}\label{native-dataframe-access}

Because the entire simulation runs in-process, the contents of a
patient's \texttt{HealthRecord} are immediately available as Python
objects without requiring serialization to disk and re-ingestion.
PySynthea exposes utilities that materialize these in-memory objects as
pandas \texttt{DataFrame} instances \citep{mckinney2010pandas}. A
typical pattern in a notebook is to generate a small cohort, immediately
materialize the conditions, encounters, observations, and medications as
separate DataFrames, and proceed with exploratory analysis or feature
engineering using familiar pandas idioms. This is the same workflow that
researchers already use with public benchmark datasets such as MIMIC-III
\citep{johnson2016mimic}, MIMIC-IV \citep{johnson2023mimic_iv}, and eICU
\citep{pollard2018eicu}, and the synthetic-data path inherits the same
ergonomics.

\subsection{ML-Ready and Notebook
Outputs}\label{ml-ready-and-notebook-outputs}

For machine learning use cases, in-memory access also makes it natural
to map records to tensors. A common requirement is to produce a sequence
of clinical events per patient --- for example, an ordered list of
(timestamp, code, value) triples --- suitable for input to a sequence
model or transformer
\citep{vaswani2017attention, rajkomar2018scalable, wornow2023foundation}.
Because the simulation has full control of event timing, the framework
can emit such sequences directly. Conversion to PyTorch
\citep{paszke2019pytorch} or TensorFlow \citep{abadi2016tensorflow}
tensors is then a simple wrapping step around the already-materialized
pandas DataFrames.

PySynthea is designed for the Jupyter notebook environment
\citep{kluyver2016jupyter}. Because installation is a single
\texttt{pip} invocation and the entire API is exposed as ordinary Python
classes, a new user can move from \texttt{pip\ install} to a generated
cohort in a small handful of cells. The \texttt{Generator} is
constructed from a \texttt{GeneratorOptions} object that is itself a
plain Python object, so users can tweak parameters interactively,
regenerate cohorts, and inspect the resulting records without leaving
the notebook. The CLI, implemented with \texttt{click}, mirrors the same
API at the shell level and shares its configuration model.

\subsection{Streaming and Batch
Generation}\label{streaming-and-batch-generation}

For larger cohorts, the engine supports both batched generation, in
which patients are produced and written to disk in chunks, and parallel
generation, in which multiple worker threads or processes simulate
independent patients concurrently. Because each patient is statistically
independent given the global seed, the simulation is embarrassingly
parallel, and PySynthea uses
\texttt{concurrent.futures.ThreadPoolExecutor} or
\texttt{ProcessPoolExecutor} to scale across cores. The \texttt{threads}
parameter in \texttt{GeneratorOptions} controls the level of
parallelism, and the exporters are designed to write incrementally so
that disk usage is bounded.

\subsection{Integration with Dask, PySpark, and
Airflow}\label{integration-with-dask-pyspark-and-airflow}

Once generation has been wrapped as a Python function, integrating it
with distributed orchestrators becomes straightforward. A Dask
\citep{rocklin2015dask} or PySpark \citep{zaharia2016spark} driver can
call the generator on each worker with a different seed and aggregate
the resulting partitioned datasets. Workflow managers such as Apache
Airflow \citep{apache_airflow} can schedule periodic regeneration of
synthetic cohorts as part of a larger pipeline, for example to refresh
test data nightly or to regenerate evaluation datasets when modules are
updated. These integrations require no special support in PySynthea
itself; they follow naturally from the fact that PySynthea is a library
that lives in the same process model as the rest of the Python data
ecosystem.

\section{Healthcare Data Standards and Export
Formats}\label{sec-formats}

Synthetic data is only as useful as the formats in which it can be
consumed. PySynthea is designed to support the major healthcare data
formats and to interoperate cleanly with general-purpose analytics
formats.

\begin{longtable}[]{@{}
  >{\raggedright\arraybackslash}p{(\linewidth - 4\tabcolsep) * \real{0.2286}}
  >{\raggedright\arraybackslash}p{(\linewidth - 4\tabcolsep) * \real{0.4000}}
  >{\raggedright\arraybackslash}p{(\linewidth - 4\tabcolsep) * \real{0.3714}}@{}}
\caption{Summary of supported export formats and typical downstream
uses.}\label{tbl-export-formats}\tabularnewline
\toprule\noalign{}
\begin{minipage}[b]{\linewidth}\raggedright
Format
\end{minipage} & \begin{minipage}[b]{\linewidth}\raggedright
Output shape
\end{minipage} & \begin{minipage}[b]{\linewidth}\raggedright
Primary use
\end{minipage} \\
\midrule\noalign{}
\endfirsthead
\toprule\noalign{}
\begin{minipage}[b]{\linewidth}\raggedright
Format
\end{minipage} & \begin{minipage}[b]{\linewidth}\raggedright
Output shape
\end{minipage} & \begin{minipage}[b]{\linewidth}\raggedright
Primary use
\end{minipage} \\
\midrule\noalign{}
\endhead
\bottomrule\noalign{}
\endlastfoot
FHIR R4 & One \texttt{Bundle} per patient, containing linked clinical
resources & Interoperability testing, FHIR server loading, SMART-on-FHIR
application development \\
CSV & One flat file per resource type, using Synthea-compatible
identifiers & pandas analysis, spreadsheet review, database bulk
loading \\
JSON & One nested document per patient & Document databases, debugging,
and inspection of complete patient histories \\
Parquet & Columnar tables derived from pandas-compatible records & Data
lakes, lakehouse storage, and distributed analytics with PySpark or
Dask \\
Relational database & Tables written through SQLAlchemy or
\texttt{pandas.to\_sql} & Research warehouses, test databases, and
downstream schema mappings \\
\end{longtable}

\subsection{FHIR R4}\label{fhir-r4}

The primary structured export format is FHIR R4
\citep{bender2013hl7, hl7_fhir_r4}. Each simulated patient is rendered
as a FHIR \texttt{Bundle} containing a \texttt{Patient} resource
together with associated resources such as \texttt{Encounter},
\texttt{Condition}, \texttt{Observation}, \texttt{MedicationRequest},
\texttt{Procedure}, \texttt{Immunization}, \texttt{AllergyIntolerance},
\texttt{CarePlan}, \texttt{Device}, \texttt{ImagingStudy}, and
\texttt{DiagnosticReport}. The exporter is configurable as either a
transaction bundle or a collection bundle, and supports optional
conformance with the US Core implementation guide. Resources reference
each other using FHIR-standard relative URIs, so the resulting bundle
can be ingested by any FHIR-conformant server, validated with standard
FHIR tooling, and used in SMART-on-FHIR applications
\citep{mandel2016smart}.

\subsection{CSV}\label{csv}

For flat-tabular consumers, PySynthea writes a directory of CSV files in
the Synthea convention, with one file per resource type (e.g.,
\texttt{patients.csv}, \texttt{encounters.csv}, \texttt{conditions.csv},
\texttt{medications.csv}, \texttt{observations.csv}). These files are
amenable to direct ingestion by pandas, by database bulk loaders, or by
spreadsheet tools, and they preserve the joining keys necessary to
reconstruct the longitudinal record.

\subsection{JSON}\label{json}

A document-oriented JSON export is also provided, in which each patient
is serialized as a single nested JSON document containing all of their
attributes, encounters, and clinical events. This format is convenient
for document databases, for NoSQL ingestion, and for hand-inspection
during development.

\subsection{Parquet and Data Lake
Integration}\label{parquet-and-data-lake-integration}

Beyond the formats inherited from upstream Synthea, the architecture
naturally supports columnar exports such as Apache Parquet
\citep{apache_parquet} via pandas or Arrow. Because the in-memory
representation is already pandas-compatible, a Parquet exporter is a
thin wrapper around \texttt{DataFrame.to\_parquet}. This makes
PySynthea-generated data directly suitable for data lakes, lakehouse
architectures, and downstream big-data analytics with PySpark.

\subsection{Database Connectors}\label{database-connectors}

For users who prefer to write directly to a relational store,
PySynthea-generated DataFrames can be written to a database through
SQLAlchemy or \texttt{pandas.to\_sql}. This allows synthetic cohorts to
be loaded into research warehouses, OMOP-style schemas (via additional
mapping), or test databases that mirror production schemas.

\section{Performance and Scalability
Considerations}\label{sec-performance}

Synthetic data generation is workload-heterogeneous: some users generate
a handful of patients in an interactive notebook, while others generate
millions for benchmarking or for pre-training experiments. PySynthea is
designed to scale across this spectrum, with several mechanisms.

\subsection{In-Process Performance}\label{in-process-performance}

The single-patient critical path is dominated by state-machine
evaluation across many timesteps. Within a single process, PySynthea
minimizes per-step overhead by pre-parsing module definitions at load
time (including converting raw code dictionaries to typed \texttt{Code}
instances once), by capping module iterations to detect cycles in
malformed modules, and by avoiding unnecessary deep copies of patient
state. Where possible, attributes are kept in dictionary structures with
O(1) access, and the simulation timestep is configurable through the
properties file (the default is one week of simulated time).

\subsection{Parallelism}\label{parallelism}

Because patients are conditionally independent given their per-patient
seeds, parallelism is straightforward. The generator exposes a
\texttt{threads} parameter that controls a pool of worker threads or
processes. For CPU-bound workloads, process-based parallelism via
\texttt{ProcessPoolExecutor} avoids the GIL and scales to the number of
physical cores. For I/O-bound workloads --- typically dominated by
export of large FHIR bundles to disk --- thread-based parallelism is
often sufficient and avoids the overhead of inter-process communication.

\subsection{Memory Usage}\label{memory-usage}

For very large runs, PySynthea avoids holding the entire population in
memory by writing each patient's exports incrementally as they finish.
The generator's bookkeeping is dominated by per-batch summary statistics
(totals of generated, living, and deceased patients, for example) and by
the loaded module definitions, which are shared across patients.

\subsection{Scaling Out}\label{scaling-out}

For scales beyond a single machine, the recommended pattern is to drive
PySynthea from a distributed orchestrator such as Dask
\citep{rocklin2015dask} or PySpark \citep{zaharia2016spark}. Each worker
invokes the generator over a disjoint seed range, and the resulting
partitioned outputs are aggregated into the target data store. Because
the generator's outputs are file-based or DataFrame-based, this
composition does not require any framework-specific support.

\subsection{Cloud-Native Execution}\label{cloud-native-execution}

The Python-native packaging also simplifies cloud deployment. A typical
pattern is to package the generator as a container image, to run it
inside a serverless or batch environment, and to write outputs directly
to cloud object storage. Because the runtime is Python, this composition
reuses the standard cloud Python toolchains, including SDKs for AWS,
Google Cloud, and Azure, without bridging through JVM-specific patterns.

\section{Example Workflows and Use Cases}\label{sec-examples}

This section illustrates how the design choices described above
translate into concrete workflows.

\subsection{Notebook-Based Cohort
Generation}\label{notebook-based-cohort-generation}

A typical workflow in a research notebook begins by importing the
package, constructing a \texttt{GeneratorOptions} object, and invoking
the generator. The user then loads the resulting FHIR bundles or CSVs
into pandas and proceeds with analysis. Because the generator is
in-process, intermediate state --- including statistics about generated,
living, and deceased patients --- is available as ordinary attributes of
the \texttt{Generator} object after the run finishes.

\begin{Shaded}
\begin{Highlighting}[]
\ImportTok{from}\NormalTok{ synthea }\ImportTok{import}\NormalTok{ Generator, GeneratorOptions}

\NormalTok{options }\OperatorTok{=}\NormalTok{ GeneratorOptions()}
\NormalTok{options.population\_size }\OperatorTok{=} \DecValTok{200}
\NormalTok{options.state }\OperatorTok{=} \StringTok{"California"}
\NormalTok{options.seed }\OperatorTok{=} \DecValTok{42}

\NormalTok{generator }\OperatorTok{=}\NormalTok{ Generator(options)}
\NormalTok{generator.run()}

\BuiltInTok{print}\NormalTok{(generator.stats)  }\CommentTok{\# e.g., \{\textquotesingle{}total\_generated\textquotesingle{}: 200, \textquotesingle{}living\textquotesingle{}: 187, \textquotesingle{}dead\textquotesingle{}: 13, ...\}}
\end{Highlighting}
\end{Shaded}

\subsection{Custom Patient
Construction}\label{custom-patient-construction}

For pedagogical or debugging purposes, individual patients can be
constructed explicitly and simulated outside of a full population run.
The example below creates a single patient with fixed attributes and
exports the resulting record to FHIR.

\begin{Shaded}
\begin{Highlighting}[]
\ImportTok{from}\NormalTok{ datetime }\ImportTok{import}\NormalTok{ datetime}
\ImportTok{from}\NormalTok{ pathlib }\ImportTok{import}\NormalTok{ Path}
\ImportTok{from}\NormalTok{ synthea.world.person }\ImportTok{import}\NormalTok{ Person}
\ImportTok{from}\NormalTok{ synthea.engine.generator }\ImportTok{import}\NormalTok{ Generator, GeneratorOptions}
\ImportTok{from}\NormalTok{ synthea.export.fhir }\ImportTok{import}\NormalTok{ FHIRExporter}

\NormalTok{person }\OperatorTok{=}\NormalTok{ Person(seed}\OperatorTok{=}\DecValTok{12345}\NormalTok{)}
\NormalTok{person.attributes.update(\{}
    \StringTok{"gender"}\NormalTok{: }\StringTok{"F"}\NormalTok{,}
    \StringTok{"birth\_date"}\NormalTok{: datetime(}\DecValTok{1980}\NormalTok{, }\DecValTok{5}\NormalTok{, }\DecValTok{15}\NormalTok{),}
    \StringTok{"first\_name"}\NormalTok{: }\StringTok{"Jane"}\NormalTok{,}
    \StringTok{"last\_name"}\NormalTok{: }\StringTok{"Doe"}\NormalTok{,}
    \StringTok{"race"}\NormalTok{: }\StringTok{"white"}\NormalTok{,}
    \StringTok{"ethnicity"}\NormalTok{: }\StringTok{"non\_hispanic"}\NormalTok{,}
\NormalTok{\})}
\NormalTok{person.init\_health\_record()}

\NormalTok{generator }\OperatorTok{=}\NormalTok{ Generator(GeneratorOptions())}
\NormalTok{generator.\_simulate\_life(person)}

\NormalTok{exporter }\OperatorTok{=}\NormalTok{ FHIRExporter(generator.config, Path(}\StringTok{"./output"}\NormalTok{))}
\NormalTok{output\_file }\OperatorTok{=}\NormalTok{ exporter.export(person, }\DecValTok{0}\NormalTok{)}
\BuiltInTok{print}\NormalTok{(}\SpecialStringTok{f"Exported to: }\SpecialCharTok{\{}\NormalTok{output\_file}\SpecialCharTok{\}}\SpecialStringTok{"}\NormalTok{)}
\end{Highlighting}
\end{Shaded}

This pattern is useful for unit testing custom modules, for teaching,
and for constructing minimal reproducible cases when debugging the
simulation engine.

\subsection{Parallel Batch Generation}\label{parallel-batch-generation}

For larger cohorts, the same API supports parallel generation. The
example below generates patients across multiple US states in parallel
using a thread pool, with deterministic seeds per state to ensure
reproducibility.

\begin{Shaded}
\begin{Highlighting}[]
\ImportTok{import}\NormalTok{ concurrent.futures}
\ImportTok{from}\NormalTok{ synthea }\ImportTok{import}\NormalTok{ Generator, GeneratorOptions}

\KeywordTok{def}\NormalTok{ generate\_batch(state, size, seed):}
\NormalTok{    options }\OperatorTok{=}\NormalTok{ GeneratorOptions()}
\NormalTok{    options.population\_size }\OperatorTok{=}\NormalTok{ size}
\NormalTok{    options.state }\OperatorTok{=}\NormalTok{ state}
\NormalTok{    options.seed }\OperatorTok{=}\NormalTok{ seed}
\NormalTok{    generator }\OperatorTok{=}\NormalTok{ Generator(options)}
\NormalTok{    generator.run()}
    \ControlFlowTok{return}\NormalTok{ generator.stats }\OperatorTok{|}\NormalTok{ \{}\StringTok{"state"}\NormalTok{: state\}}

\NormalTok{states }\OperatorTok{=}\NormalTok{ [}\StringTok{"California"}\NormalTok{, }\StringTok{"Texas"}\NormalTok{, }\StringTok{"New York"}\NormalTok{, }\StringTok{"Florida"}\NormalTok{]}
\ControlFlowTok{with}\NormalTok{ concurrent.futures.ThreadPoolExecutor(max\_workers}\OperatorTok{=}\DecValTok{4}\NormalTok{) }\ImportTok{as}\NormalTok{ ex:}
\NormalTok{    results }\OperatorTok{=} \BuiltInTok{list}\NormalTok{(ex.}\BuiltInTok{map}\NormalTok{(}
        \KeywordTok{lambda}\NormalTok{ i\_state: generate\_batch(i\_state[}\DecValTok{1}\NormalTok{], }\DecValTok{100}\NormalTok{, i\_state[}\DecValTok{0}\NormalTok{] }\OperatorTok{*} \DecValTok{1000}\NormalTok{),}
        \BuiltInTok{enumerate}\NormalTok{(states),}
\NormalTok{    ))}
\end{Highlighting}
\end{Shaded}

\subsection{ML Benchmarking and Medical
NLP}\label{ml-benchmarking-and-medical-nlp}

Because synthetic cohorts can be regenerated arbitrarily, they are well
suited for benchmarking machine learning pipelines under controlled
conditions. A researcher can fix a seed, generate a cohort, train a
model, and rerun the same experiment elsewhere with bit-exact data.
Cohorts can be parametrically varied --- for example, by changing
disease prevalence or the distribution of ages --- to test model
robustness. Medical NLP pipelines can be primed on synthetic discharge
summaries or structured records before being applied to real, smaller,
protected corpora \citep{wornow2023foundation, rajkomar2018scalable}.

\subsection{Federated Learning
Simulation}\label{federated-learning-simulation}

For federated learning experiments \citep{mcmahan2017communication},
synthetic cohorts can stand in for the per-institution shards of a
federation. Each shard can be parameterized differently to reflect
institutional heterogeneity, and the resulting setup can be used to
study algorithmic properties of federated training without requiring
access to data from any real institution.

\subsection{Education and Healthcare Interoperability
Testing}\label{education-and-healthcare-interoperability-testing}

In educational settings, PySynthea allows instructors to distribute a
single notebook that produces a known, reproducible cohort, and to build
entire lecture sequences around it. In software development settings,
the same generator can produce a controlled set of FHIR bundles for
end-to-end interoperability tests, replacing brittle hand-crafted
fixtures and ensuring that test data covers a wide range of realistic
clinical scenarios.

\section{Discussion}\label{sec-discussion}

The design of PySynthea trades certain properties of the original
Java-based Synthea against others that are more important in modern
Python-based workflows. We discuss these tradeoffs explicitly.

\subsection{Fidelity Versus
Accessibility}\label{fidelity-versus-accessibility}

The first tradeoff is between absolute fidelity to the upstream Synthea
reference implementation and the ergonomic accessibility of a clean
Python rewrite. Because PySynthea consumes the same GMF JSON modules
unchanged, it preserves the bulk of the clinical content that gives
Synthea its value. However, the engine itself is an independent
reimplementation, and small semantic differences --- for example, in how
rare edge cases in transitions or guards are evaluated --- are possible.
We treat the upstream Java implementation as the semantic reference and
aim to converge with it on shared modules, with explicit divergence
tracked in the project's documentation.

\subsection{Statistical Realism}\label{statistical-realism}

Like upstream Synthea, PySynthea generates patients whose statistical
properties are calibrated against external references (census
demographics, growth charts, cost data, immunization schedules) but
whose joint statistical structure does not faithfully reproduce that of
any particular real cohort. Synthetic Synthea-style data is therefore
useful for structural and pipeline-level applications --- schema
validation, interoperability testing, model engineering, education ---
and for many forms of methodological experimentation, but it should not
be confused with a generative model trained on real data, such as a
generative adversarial network or a diffusion model applied to EHRs
\citep{choi2017medgan, yan2022multifaceted}. We see these two classes of
synthetic data as complementary rather than competing: Synthea-style
data provides clinically structured longitudinal records with full
provenance, while learned generative models provide higher fidelity to
specific distributions at the cost of interpretability and standards
compliance.

\subsection{Python Performance
Considerations}\label{python-performance-considerations}

A common concern about Python-native reimplementations of CPU-intensive
workloads is performance. In our experience, the per-patient simulation
cost is dominated by state-machine evaluation and by serialization, both
of which can be optimized within Python through pre-parsing, simple data
structures, and efficient JSON or Parquet writers. For workloads beyond
what a single Python process can sustain, the recommended path is
horizontal scaling via processes, Dask, or PySpark, rather than vertical
optimization of a single process. Future versions may use Cython, Numba,
or Rust-based extensions to accelerate the inner loop of state-machine
evaluation if profiling indicates that this is the binding constraint.

\subsection{Long-Term Maintenance}\label{long-term-maintenance}

Reimplementing a substantial open-source project always raises questions
about long-term maintenance. PySynthea is explicitly designed to consume
upstream Synthea's module library, so improvements to clinical content
there flow into PySynthea without code-level synchronization. The engine
itself is more compact than the full Java codebase, and is organized so
that contributors familiar with Python and basic state-machine concepts
can extend it without needing to learn a new toolchain.

\subsection{Compatibility with Upstream
Synthea}\label{compatibility-with-upstream-synthea}

A core design constraint is that PySynthea should remain a ``good
citizen'' within the broader Synthea ecosystem. This shows up in three
places. First, PySynthea consumes the same GMF JSON modules, so the
shared community library of disease content remains the single source of
truth. Second, PySynthea produces FHIR bundles, CSVs, and JSON documents
that are compatible at the schema level with upstream Synthea, so
existing consumers --- FHIR servers, validation pipelines, analytics
tooling --- can ingest PySynthea outputs without modification. Third,
PySynthea adopts the same configuration model and many of the same CLI
options, so users moving between the two implementations encounter a
familiar interface.

\section{Future Work}\label{sec-future}

PySynthea opens several lines of future research and engineering work,
each of which builds on the Python-native foundation described above.

\emph{GPU acceleration.} The inner loop of state-machine evaluation,
particularly when run over millions of patients, is a candidate for GPU
acceleration. While the current implementation uses CPU-based
parallelism, longer-term work may explore vectorized or batched
simulation strategies that exploit GPUs through CUDA, JAX, or PyTorch,
especially for the subset of states whose semantics are amenable to
batched evaluation.

\emph{LLM-generated and LLM-assisted disease modules.} Large language
models have demonstrated significant capability in authoring structured
clinical content \citep{wolf2020transformers, wornow2023foundation}. A
natural extension is to use LLMs to draft GMF modules from clinical
guidelines or literature, with human-in-the-loop review. PySynthea's
Python-native nature makes it straightforward to embed such workflows in
the same environment used to train or evaluate the LLMs themselves.

\emph{Multi-agent patient simulation.} Synthea's current model treats
patients as independent, but realistic healthcare populations include
interaction effects --- outbreaks, household transmission, healthcare
system congestion --- that are best modeled with multi-agent simulation.
Extending PySynthea with multi-agent dynamics, where individual patient
simulations are coupled through shared environmental state, opens new
research possibilities in infectious disease modeling and healthcare
operations research.

\emph{Reinforcement learning environments.} The discrete-time,
state-driven nature of Synthea makes it a natural substrate for
reinforcement learning environments grounded in clinical decision
making. A PySynthea-backed environment that exposes patient state to an
agent and accepts treatment actions could serve as a benchmark for
offline and online RL in clinical care, while keeping the underlying
data fully synthetic.

\emph{Differential privacy and privacy-preserving generation.} Although
Synthea-style data does not derive from real patients and therefore does
not require privacy guarantees in the traditional sense, the framework
can serve as a useful testbed for differentially private machine
learning algorithms \citep{dwork2006differential} and for evaluating
membership inference risk in models trained on simulated cohorts.

\emph{Synthetic imaging integration.} Modern healthcare data is
increasingly multimodal. Integrating PySynthea with synthetic imaging
pipelines --- for example, generative models for chest X-rays or
pathology images linked to simulated clinical histories --- would enable
end-to-end multimodal benchmarks that combine structured EHR data with
synthetic medical images.

\section{Conclusion}\label{sec-conclusion}

Synthetic healthcare data is an increasingly important resource for
research, education, software engineering, and applied machine learning
in medicine. By generating realistic longitudinal records without
disclosing information about real individuals, it sidesteps the
regulatory and operational constraints that limit access to real EHRs
and provides a substrate for reproducible, shareable, and scalable
experimentation. Synthea has established itself as a leading tool in
this space, with a state-machine-based simulation model, a rich library
of disease modules, and broad support for healthcare interoperability
standards \citep{walonoski2018synthea, walonoski2020covid}.

The friction in current workflows, however, is no longer about the
quality of the synthetic data itself. It is about the alignment between
the data generator and the dominant data science ecosystem. Python is
now the de facto standard for data analysis, machine learning, and AI in
healthcare and elsewhere, and tools that require external runtimes or
file-based bridges into Python pay a real cost in adoption, in
reproducibility, and in the velocity of research.

PySynthea addresses this gap by reimplementing the Synthea engine as a
pure-Python, pip-installable framework that preserves the Generic Module
Framework, supports standards-conformant export formats, and integrates
naturally with pandas, NumPy, PyTorch, Dask, PySpark, Airflow, and
Jupyter. It is designed to be accessible from the first line of a
notebook, reproducible across runs, extensible by researchers in their
native language, and scalable from interactive sessions to distributed
batch jobs. By lowering the cost of adopting synthetic healthcare data
and by aligning the generator with the rest of the modern data science
toolchain, PySynthea aims to broaden participation in
synthetic-data-driven research and to accelerate the responsible
development of machine learning systems for medicine.

\section*{Acknowledgements}\label{acknowledgements}
\addcontentsline{toc}{section}{Acknowledgements}

The authors thank the TietAI clinical team for feedback on governance
requirements and the Hydra Platform engineering team for integration
support.

\bibliographystyle{unsrt}
\bibliography{refs.bib}

\end{document}